\documentclass[runningheads]{llncs}
\usepackage{graphicx}
\usepackage{multirow}
\usepackage{booktabs}
\usepackage{xcolor}
\begin{document}
\title{Distributed Intrusion Detection in Dynamic Networks of UAVs using Few-Shot Federated Learning}
%
%

\author{Ozlem Ceviz\inst{1}\orcidID{0000-0002-8610-4008} \and Sevil Sen\inst{1}\orcidID{1111-2222-3333-4444} \and
Pinar Sadioglu\inst{1}\orcidID{0000-0003-2287-5177}}
\authorrunning{O. Ceviz et al.}
%
\institute{WISE Lab., Department of Computer Engineering, Hacettepe University, Ankara, Turkey\\ 
\email{ozlemceviz@hacettepe.edu.tr} \\
\email{ssen@cs.hacettepe.edu.tr}\\
\email{pinar.sadioglu@gmail.com}}

%
\maketitle              
\begin{abstract}
Flying Ad Hoc Networks (FANETs), which primarily interconnect Unmanned Aerial Vehicles (UAVs), present distinctive security challenges due to their distributed and dynamic characteristics, necessitating tailored security solutions. Intrusion detection in FANETs is particularly challenging due to communication costs, and privacy concerns. While Federated Learning (FL) holds promise for intrusion detection in FANETs with its cooperative and decentralized model training, it also faces drawbacks such as large data requirements, power consumption, and time constraints. Moreover, the high speeds of nodes in dynamic networks like FANETs may disrupt communication among Intrusion Detection Systems (IDS). In response, our study explores the use of few-shot learning (FSL) to effectively reduce the data  required for intrusion detection in FANETs. The proposed approach called Few-shot Federated Learning-based IDS (FSFL-IDS) merges FL and FSL to tackle intrusion detection challenges such as privacy, power constraints, communication costs, and lossy links, demonstrating its effectiveness in identifying routing attacks in dynamic FANETs.This approach reduces both the local models and the global model's training time and sample size, offering insights into reduced computation and communication costs and extended battery life. Furthermore, by employing FSL, which requires less data for training, IDS could be less affected by lossy links in FANETs.

\keywords{FANETs \and UAV \and dynamic networks \and intrusion detection \and few-shot federated learning.}
\end{abstract}

\section{Introduction}
Unmanned Aerial Vehicles (UAVs), also known as drones, have gained popularity in recent years due to advancements in technology. With their versatile capabilities for autonomous operation by a computer or remote control, UAVs have found applications in diverse fields such as military missions, search and rescue operations, target tracking, and agricultural tasks. This widespread adoption is evident in the United States, where approximately 900,000 UAVs have been registered as of 2022 \cite{satell2022drone}. As the number of UAVs continues to rise, the need for effective coordination in shared airspace becomes paramount. This challenge is met through the integration of Flying Ad Hoc Networks (FANETs) \cite{bekmezci2013flying}, which plays a pivotal role in ensuring organized communication among UAVs or drones. However, dynamic and decentralized characteristics of FANETs cause significant security challenges, requiring a robust Intrusion Detection Systems (IDSs).

In recent years, there has been a notable emphasis on IDSs within the context of FANETs, often utilizing Machine Learning (ML) techniques to enhance detection capabilities \cite{zhai2023etd}\cite{chulerttiyawong2023sybil}. The integration of ML, and Deep Neural Networks (DNN) has shown promising outcomes in identifying abnormal network patterns \cite{nayfeh2023machine}\cite{Ouiazzane2020}. However, challenges such as communication costs and privacy concerns have emerged as significant obstacles. 
These ML-based approaches often struggle to address the potential exposure of sensitive information and integrate diverse threats in real-time, which may affect the detection of unknown attacks \cite{hu2023privacy}.
Moreover, the distributed architecture of FANETs amplifies these limitations, raising questions about the suitability of current methods for such environments.
The substantial need for a large amount of data, along with significant costs in terms of power consumption and time \cite{vinayakumar2017applying,lu2021cognitive,lu2023few}, is another notable drawback of employing IDS for FANETs with traditional algorithms. This challenge becomes even more apparent when considering the frequent connectivity disruptions caused by the dynamic nature of FANETs, resulting in significant data losses during the data collection process. Moreover, like zero-day attacks or new types of threats, there is often insufficient time to gather a significant number of malicious samples for learning from these latest attacks  \cite{ma2023adversarial}. Effectively addressing this issue necessitates a pivotal shift towards conducting training and decision-making with minimal data.

Federated Learning (FL) represents a promising paradigm for IDSs, wherein numerous decentralized UAVs possess the capacity to independently train localized models \cite{agrawal2022federated}. These UAVs exclusively transmit the derived model parameters to a centralized server, thereby avoiding the sharing of sensitive data, for subsequent aggregation. This approach enables rapid decision-making in detecting attacks without the delay of collecting data from UAVs, addressing privacy concerns, and enhancing the system's effectiveness in responding to security threats \cite{ceviz2023novel}. Another method worth considering is Few-Shot Learning (FSL), which specifically addresses challenges related to learning from a limited number of samples. FSL aims to overcome this limitation by extracting valuable insights from only a small number of samples. This method becomes especially valuable in scenarios where computational and storage efficiency is critical, such as for resource-constrained UAVs operating on small batteries.

The combination of FL and FSL is a promising avenue in research, with existing studies validating its feasibility \cite{hu2023privacy}. This study contributes to this evolving field by introducing a Few-shot Federated Learning-based Intrusion Detection System (FSFL-IDS) in FANETs. FSFL-IDS effectively addresses key challenges in intrusion detection, including privacy concerns, the substantial computational power consumption resulting from handling large datasets, and the communication cost in decentralized architectures. It efficiently minimizes both the locals' and global's models training time and sample size. The contributions of this study can be summarized as follows:

\begin{itemize}
    \item A novel intrusion detection approach for FANETs called FSFL-IDS that integrates FL and FSL techniques to utilize the unique advantages of each method is introduced. This approach addresses the challenge of intrusion detection in FANETs when facing insufficient sample sizes due to node constraints and packet losses. In addition to its effectiveness, it offers advantages such as reduced communication, computational and storage costs, extended battery life, and enhanced privacy.
    \item FSFL-IDS is thoroughly tested using a FANET dataset containing diverse attack types (e.g., sinkhole, blackhole, flooding). Evaluations cover attacker percentages from 5\% to 25\%, along with attack-free scenarios, mirroring real-world FANET simulations. The results show that the proposed approach achieves comparable outcomes to FL-based IDS, despite utilizing substantially less data (10\%).
    
    \item FSFL-IDS is compared with a traditional central IDS, where local data is periodically shared, potentially raising privacy and communication concerns. The comparison shows that FSFL-IDS, with local UAVs learning independently and benefiting from global model updates, achieves comparable results even with 20-shots.

    \item The proposed approach shows FSFL-IDS's adaptability and suitability for dynamic, resource-constrained networks. Its versatility suggests potential application in other dynamic network types in the future.

\end{itemize}
\section{Related Work}

The related studies in the literature will be covered under two topics: FL-based approaches for FANETs, and FSL techniques used for intrusion detection.

\subsection{Federated Learning-based IDS for FANETs}

While traditional centralized techniques can effectively detect malicious activities and offer promising results for enhancing FANET security, they often overlook privacy concerns and issues arising from communication over lossy links and associated communication costs. Although federated learning has high potential for addressing such issues, there is a limited number of studies \cite{mowla2019federated,mowla2020afrl,da2023anomaly,ceviz2023novel} in the literature.

In \cite{mowla2019federated}, a FL-based approach was employed to detect jamming attacks, using datasets for FANETs and Vehicular Ad hoc Networks (VANETs). The CRAWDAD VANET dataset \cite{punal2014crawdad} 
was adapted to address the unique unbalanced data characteristics of FANETs. The implementation included a UAV selection method based on the Dempster-Shafer theory, enhancing FL efficiency. It achieved approximately 82\% accuracy for CRAWDAD and 89.5\% for the FANET dataset, showcasing superior performance compared to traditional solutions. In their subsequent extended study \cite{mowla2020afrl}, a reinforcement FL approach was introduced to identify a defense strategy in unknown environments. It devised alternative routes, strategically avoiding areas susceptible to jamming attacks through spatial retreat. In \cite{da2023anomaly}, a FL-based approach is proposed to detect GPS jamming and spoofing attacks using the UAV Attack Dataset \cite{whelan2020novelty}. It is shown that the FedAvg algorithm stands out for its robust performance, achieving an F1-score of 0.887.

In a recent study \cite{ceviz2023novel}, a Federated Learning-based Intrusion Detection System (FL-IDS) was introduced for detecting routing attacks in FANETs. Initially, a dataset comprising 50 nodes with 3D movements was created for simulating realistic UAV applications. FL-IDS was compared with central and local IDSs and shown to perform close to the central IDS in most experiments, highlighting its potential for the distributed architecture of FANETs. Moreover, it utilized the Bias Towards Specific Clients (BTSC) approach to enhance detection performance.

\subsection{Few-shot Learning-based IDS}

The challenge of intrusion detection with few attack samples, especially on resource-constrained devices, drives researchers to explore FSL applications \cite{gamal2021few,ma2023adversarial,lu2023few}. Another motivation is the inability of many IDS to detect new attack types due to limited data availability for such instances. In \cite{gamal2021few}, an IDS is proposed to detect anomalies in edge networks. The proposed system contains two phases: feature selection and decision engine. The decision engine, using FSL techniques with CNN, is designed for detecting new attack types at the network edge. Training involves benchmark datasets from UNSW-NB15 \cite{moustafa2015unsw} and Bot-IoT \cite{koroniotis2019towards}. 

The FSL technique, coupled with domain adaptation, has emerged as a solution for detecting malicious traffic in scenarios with limited data in IoT \cite{ma2023adversarial}. Domain adaptation, a technique within transfer learning, plays a crucial role in aligning the feature distributions of different domains, especially in situations where labeled data is scarce in the target domain. The approach involves initially training ML-based detection models with a substantial dataset from a source network. Subsequently, the models are fine-tuned using only a few shots from the target network. Gathering a minimal number of few-shot target domain samples is both practical and feasible, offering a more cost-effective alternative compared to collecting a large number of unlabeled samples.

Another FSL-based technique, utilizing CNN and meta-learning, was proposed in \cite{lu2023few}. The study employed a comprehensive dataset formed by merging five distinct datasets. The N-way K-shot selection was implemented, with cases including 5-way 1-shot, 5-way 5-shot, and 5-way 10-shot. Here, "N" signifies the number of classes, while "K" denotes the number of examples or "shots" available for each class. Their best outcome achieved an accuracy of 92.19\% in the case of the 5-way 10-shot setting. 

While FSL has been applied in various domains like IoT and edge networks recently, to the best of our knowledge, it has not been applied to FANETs. This is noteworthy because FANETs are dynamic networks that frequently experience connectivity disruptions and potential data losses due to the high speed of UAVs. Additionally, energy consumption remains a critical design consideration, especially with the use of mini UAVs powered by low-capacity batteries. Node constraints emphasizes the urgent requirement for innovative lightweight solutions. Hence, FSL is a technique worth investigating for intrusion detection in FANETs. Consequently, these unique dynamics differentiate them from the environments typically examined in previous research. Moreover, it is used together with federated learning in order to provide privacy in this study.

Related studies in the literature are  summarized in Table \ref{tab:rw}. To summarize, the majority of studies on intrusion detection in FANETs primarily focus on traditional ML-based methods, overlooking the unique distributed characteristics of FANETs. While a few studies have proposed approaches based on federated learning, which is well-suited for the distributed nature of FANETs, it is important to note that the associated communication costs tend to increase with a higher number of clients or the implementation of more complex IDS architectures. Moreover, these methods require large datasets, which can be burdensome for resource-constrained nodes. However, acquiring real-world data can be challenging and costly, highlighting the need to address threats with limited data. Some studies address this by exclusively applying FSL methods in various domains such as edge networks \cite{gamal2021few} and IoT \cite{ma2023adversarial}\cite{lu2023few}. However, there is currently a lack of research on intrusion detection for FANETs using few-shot learning in the literature.

In light of the advantages offered by federated and few-shot learning, IDS research has moved towards to use the combination of both techniques \cite{hu2023privacy}. This current study introduces a novel application of few-shot learning integrated with federated learning to a specially designed dataset for FANETs.  The proposed approach called Few-Shot Federated Learning-based IDS (FSFL-IDS), contributes to the development of a suitable detection system for the unique network environment of FANETs. The proposed method aligns well with the distributed structure of FANETs, offering privacy benefits and reducing computation and communication costs for resource-constrained devices. As emphasized earlier, the necessity to train IDS with minimal data, particularly in the face of the risk of dropped data packets in the challenging conditions of FANETs, underscores the critical need for swift decision-making and minimizing potential damage from network attacks.

\begin{table*}[t]
\caption{Outline of Related Studies}
\label{tab:rw}
\resizebox{\textwidth}{!}{
\begin{tabular}{|c|l|l|l|l|l|}
\hline
\multicolumn{1}{|l|}{Method} &
  Study &
  Year &
  Environment &
  Dataset &
  Architecture \\ \hline

\multirow{4}{*}{FL} &
  \cite{mowla2019federated} &
  2019 &
  FANETs &
  FANET Dataset, CRAWDAD VANET dataset \cite{punal2014crawdad} &
  Distributed Training \\ \cline{2-6} 
 &
  \cite{mowla2020afrl}, &
  2020 &
  FANETs &
 FANET Dataset, CRAWDAD VANET dataset \cite{punal2014crawdad} &
  Distributed Training \\ \cline{2-6} 
 &
  \cite{da2023anomaly} &
  2023 &
  UAV swarm &
  UAV Attack Dataset \cite{whelan2020novelty} &
  Distributed Training \\ \cline{2-6} 
 &
  \cite{ceviz2023novel} &
  2023 &
  FANETs &
  FANET Dataset &
  Distributed Training \\ \hline
\multirow{4}{*}{FSL} &
  \cite{gamal2021few} &
  2021 &
  IoT &
  UNSW-NB15 \cite{moustafa2015unsw}, Bot-IoT \cite{koroniotis2019towards} &
  Central Training \\ \cline{2-6} 
 &
  \cite{ma2023adversarial} &
  2023 &
  IoT &
  UNSW-NB15 \cite{moustafa2015unsw}, TON\_IoT \cite{alsaedi2020ton_iot}, a IoT dataset &
  Central Training \\ \cline{2-6} 
 &
  \multirow{2}{*}{\cite{lu2023few}} &
  \multicolumn{1}{c|}{\multirow{2}{*}{2023}} &
  \multirow{2}{*}{IoT} &
  CIC- DDoS2019 \cite{zhang2019intrusion}, CIC-IDS2017 \cite{panigrahi2018detailed}, &
  \multirow{2}{*}{Central Training} \\ \cline{5-5}
 &
   &
  \multicolumn{1}{c|}{} &
   &
  CSE-CIC-IDS2018 \cite{sharafaldin2018toward}, NSL-KDD \cite{5356528}, and UNSW-NB15 \cite{moustafa2015unsw} &
   \\ \hline
FSL + FL &
  Our study &
  2024 &
  FANETs &
  FANET Dataset &
  Distributed Training \\ \hline
\end{tabular}}
\end{table*}

\section{BACKGROUND}\label{metodology}

This section offers background on ML techniques, namely FL and FSL, alongside the Hyperband algorithm used for hyperparameter optimization in this study.

\subsection{Federated Learning }

Federated Learning (FL) \cite{zhang2021survey} enables model training across decentralized devices holding local data samples, without exchanging them. This approach is particularly useful in privacy-sensitive scenarios where the data cannot be easily shared due to legal, security, or confidentiality concerns. The FL process involves several key steps. Initially, a global model is defined and distributed to all participating devices. Each device then independently computes model updates based on its local data. 
Rather than sending raw data to a central server, only model updates in the form of weights are transmitted. These updates are aggregated at a central server to refine the global model based on collective knowledge from all devices. This iterative collaboration enhances the global model's performance over time while preserving individual dataset privacy. 

FL emerges as a fitting and innovative approach to address challenges posed by the decentralized and dynamic networks. FANETs operate in a distributed fashion, communicating wirelessly while constantly changing their positions. In this context, traditional centralized ML models may face difficulties due to privacy concern and communication constraints to dynamic network conditions. However, FL, with its decentralized training mechanism, can aligns well with the characteristics of FANETs. FL is especially attractive for IDSs aiming to achieve efficiency, privacy preservation, and decentralization in FANETs.

\subsection{Few-shot Learning}

Few-shot learning (FSL) \cite{wang2020generalizing} focuses on using small datasets to train models, making it particularly useful for intrusion detection. In contrast to traditional  models, FSL does not directly predict the class of an input sample, but puts an emphasis on calculating the distance between input samples within their feature representations. These feature representations are learned during the training process and are designed to transform input samples into a high-dimensional feature space where similar samples are brought closer, and dissimilar samples are pushed farther apart. Additionally, even when the training data is limited, the model's parameters are optimized during the training process to ensure the effectiveness of the learned features in distinguishing between different classes.

In FSL, the distance between feature of two input samples, $x_i$ and $x_j$, is determined and computed using pairwise Euclidean distance. This can be expressed as follows:
\begin{equation}
    D(f(x_i), f(x_j)) = \| f(x_i) - f(x_j) \|^2
\end{equation}

Ultimately, the FSL produces its output by utilizing the fully connected layer and sigmoid layer. This process can be described through the equation below: 
\begin{equation}
   P(x_i, x_j) = {Sigmoid}({FCNN}(D(f(x_i), f(x_j))))
\end{equation}

where Sigmoid(*) denotes the Sigmoid function, FCNN(*) signifies the fully connected layer function for CNN, and \(P(x_i, x_j)\) denotes the probability of whether samples \(x_i\) and \(x_j\) belong to the same class.

When we include K samples of each of the N possible classes in the training data for learning a classification model in FSL, it is called N-way K-shot learning. During training, K labeled samples are selected from each class to form a sample set S, denoted as \(\{(x_i, y_i)\}_{i=1}^m\), where \(m\) represents the total number of samples, calculated as K multiplied by the number of classes. In a distributed system, this should be computed separately for each UAV (i.e., N * K * number of UAVs). This process equips the model with the ability to quickly adapt to new, unseen tasks during testing.  

ML-based methods often require large datasets to learn system patterns, necessitating significant effort and expense for data gathering and labeling to establish a reliable dataset. With network data volumes reaching terabyte levels within a single day, the manual analysis and labeling of such extensive data become impractical, heavily reliant on expert knowledge \cite{yang2022fs}. As a result, allocating sufficient time and resources for the collection, analysis, and labeling of the latest attack samples for intrusion detection models becomes increasingly infeasible.

In scenarios where acquiring large training data is impractical or costly, FSL proves highly advantageous for FANETs. Operating in resource-constrained conditions, FANET devices have limited computational capabilities and battery life. FSL, focusing on model training with small datasets, aligns seamlessly with these constraints. Moreover, FANETs' dynamic nature often results in frequent connectivity disruptions, leading to significant data loss during collection. Hence, minimizing data usage during training/decision-making is crucial to effectively address this challenge, the primary aim of this study.

\subsection{Hyperband Algorithm}

Hyperband \cite{li2018hyperband} is a hyperparameter optimization algorithm aimed at efficiently discovering optimal parameters, crucial for a model's performance. The process can be computationally expensive, involving training and evaluating numerous models with different hyperparameter configurations. Hyperband addresses this challenge by combining random sampling with a bandit strategy, efficiently allocating computational resources. It runs multiple configurations in parallel, utilizing intermediate results to prioritize resource allocation. This combination of random sampling, successive halving, and aggressive elimination renders Hyperband a powerful and resource-efficient algorithm for hyperparameter optimization, outperforming traditional methods like random search and grid search in both speed and effectiveness \cite{li2018hyperband}.

\section{THE PROPOSED APPROACH}

This section outlines the proposed approach, Few-shot Federated Learning-based IDS (FSFL-IDS), along with the implemented attacks and experimental settings.

\subsection{Few-shot Federated Learning-based IDS}

\begin{figure}[h]
\centering
\includegraphics[width=\textwidth]{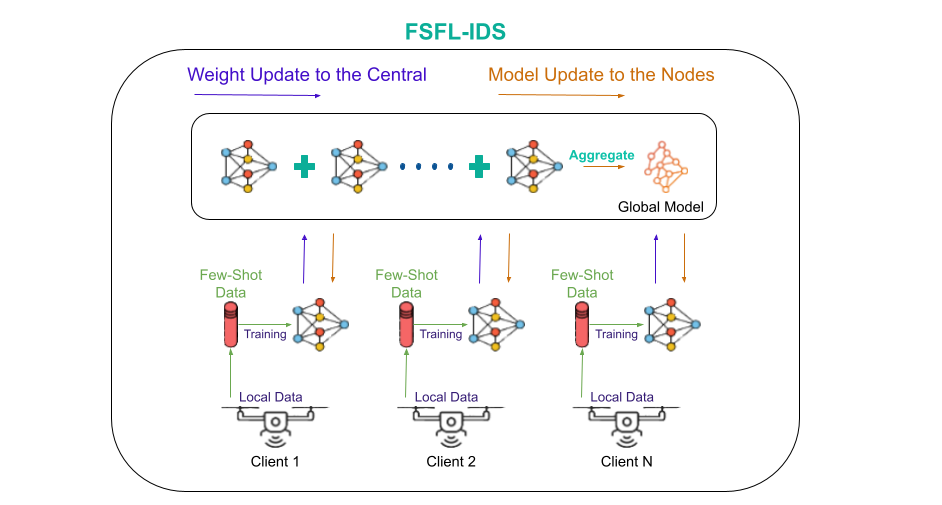}
\caption{Few-Shot Federated Learning-based IDS (FSFL-IDS)}
\label{FSFL-IDS}
\end{figure}

The proposed FSFL-IDS approach, illustrated in Figure \ref{FSFL-IDS}, combines FL and FSL techniques. Each network node actively engages in intrusion detection, as shown. This method allows distributed UAVs to collectively train a model without sharing their local data, using FL. Moreover, it employs FSL, utilizing a limited number of local data samples for training.

Each model sends its weights to a central server, where the Federated Averaging (FedAvg) methodology is employed. This technique aggregates the weights contributed by each UAV to create a global model. The global model then distributes its updates to each UAV, ensuring the privacy of UAV data while facilitating collaborative model training across distributed devices. Furthermore, the global model continues its training process by incorporating periodic updates obtained from the few-shot data available at each UAV. This process is limited to 10 epochs in this study, aiming to minimize communication costs by utilizing a smaller number of epochs.

\begin{table}[h!]
\caption{Parameters of FSFL-IDS}
\centering
\label{table:parameters}
\resizebox{.6\textwidth}{!}{\begin{tabular}{l l l } 
\toprule
 Algorithms &Parameter& Value \\ 
  \midrule

 DNN & Number of neurons &10 \\ 
 &Number of hidden layers&2\\
 \hline
 CNN & Number of nodes&6\\
 &Number of hidden layers&1\\
 &Convolutional layers& 1 Conv1D\\
 &Pooling layers&1 MaxPool1D\\
 &Dropout& 0.2 \\
 &Kernel size&3\\
 &Filters&9\\
 \hline
 Used in both & Optimizer&Adam\\
 &Learning rate &0.001 - 0.01\\
 &Loss function &binary\_crossentropy\\
 &Batch size &5\\
 &Activation function &ReLu\\
 &Classification function &sigmoid\\
 &Number of local epoch&1\\
 &Number of global epoch&10\\

\bottomrule
\end{tabular}}
\end{table}

As stated above, FSFL-IDS employs DNN and CNN algorithms to train a global model separately. These algorithms involve several parameters that require fine-tuning to achieve the best performance, and the Hyperband algorithm is employed to determine the most effective parameters for each models, which are listed in Table \ref{table:parameters}. FSFL-IDS is trained using the dataset, the details of which will be explained in the subsequent subsection. Although this dataset was previously generated with each node collecting local data every 5 seconds in our simulation \cite{ceviz2023survey}, only 10\% of this data is utilized in the first FSFL-IDS experiment (each UAV has 36 samples).  The data is divided into training (80) and testing sets (20). 

In order to enhance the model’s performance and convergence,
the StandardScale method is employed for both the training and
test sets. This method involves centering and standardizing the
input features, ensuring a mean of 0 and a standard deviation of
1. This standardized representation not only facilitates improved
algorithm performance but also maintains interpretability, enabling
fair comparisons across different features. 
FSFL-IDS explores the problem within N-way K-shot settings, wherein a classifier model undergoes training through CNN or DNN training. This approach results in a total of N*K data points for N-way classification in each UAV. Here, the problem at hand is 2-way, a binary classification problem distinguishing malicious traffic from benign traffic. Furthermore, in this context, three different sample sizes, namely 36, 20, and 10 samples, are utilized for each UAV from the two respective classes in the experiments.

\subsection{Experimental Settings}

In this study, three routing attacks, namely sinkhole, blackhole, and ad hoc flooding against the Ad hoc On-Demand Distance Vector Routing (AODV) routing protocol within FANETs, are implemented using the Ns-3 simulator \cite{Ns-3}. Each network consists of 50 mobile nodes and one immobile Ground Base Station (GBS) node, which is placed at the center of the simulation area. With this setup, our scenario emulates both the Air-to-Air (A2A) communication channel and the communication link between UAVs and Ground Base Stations (GBS) as Air-to-Ground (A2G) channel communication. The 3D natural flight of UAVs is simulated using the 3D Gauss Markov (GM) mobility model, which incorporates realistic movement patterns.  Attacker nodes are randomly chosen among the non-source or non-destination nodes. For each attacker ratio (5\% to 25\%), the same nodes are consistently used in each attack type. The network simulation parameters are  outlined in Table \ref{table:simsetup}. Further details on parameter selection can be found in our previous study \cite{ceviz2023survey}. 

\begin{table}[h!]
\caption{Network Simulation Parameters}
\label{table:simsetup}
\centering
\resizebox{.6\columnwidth}{!}{\begin{tabular}{|c |c |} 
 \hline
 Parameters & Values \\ [0.5ex] 
 \hline
 Simulation area &  12000 m x 12000 m x 300 m\\
\hline\hline
 Simulation time & 1800 seconds\\
 \hline
 \# of nodes & 50\\
\hline
Average speed & 100 m/s \\
\hline
Transmission range & 250 m\\
\hline
 Routing protocol & AODV \\ 
 \hline
 MAC protocol & IEEE 802.11b\\
 \hline
Traffic type & UDP with 10 connections \\
\hline
Packet size & 512 bytes\\
\hline
Packet rate & 1/s \\
\hline
Bandwidth & 11 Mbps \\
\hline
Attacker ratio & no attack, 5\%, 10\%, 15\%, 20\%, 25\%\\
\hline
Mobility model & 3D GM \\
\hline
Bounds for GM Mobility & X: [0; 12000],
Y: [0; 12000], Z: [0; 300]\\
\hline
$\alpha$ for mobility model & [0.25-0.7]\\ [1ex]
 \hline
\end{tabular}}
\end{table}

\textbf{\emph{Ad hoc On-Demand Distance Vector (AODV)}} is a routing protocol designed for Mobile Ad hoc Networks (MANETs). It operates reactively, establishing routes only when needed for data transmission. 
AODV is known for its low overhead and adaptability, making it suitable for use in networks with highly dynamic topologies such as FANETs.

AODV initiates route discovery by broadcasting Route Request (RREQ) packets, which propagate through the network until reaching the destination or a node with a fresh route. Upon receiving a RREQ, a unicast Route Reply (RREP) packet is sent back to the source, facilitating route establishment based on the shortest path and highest sequence number, indicating the freshest route. The protocol includes a route maintenance mechanism using Route Error (RERR) messages to handle link failures. When a link failure occurs, affected routes are invalidated, and new route discoveries are initiated.

\textbf{\emph{Routing Attacks:}}
There attacks are implemented in this study. In \emph{sinkhole attack}, upon receiving an RREQ packet from the source node, the attacker generates a fake RREP packet with an increased destination sequence number, falsely claiming to be one hop away from the destination. This ensures that network traffic is directed through the attacker nodes. This attack is commonly executed as an initial step for subsequent malicious actions, such as dropping/modifying packets. 

The sinkhole attack is integrated with a dropping attack, and a \emph{blackhole attack} is implemented as the second scenario. Initially, the attacker attracts traffic to itself, as mentioned in the sinkhole attack strategy. Subsequently, it drops every data packet that it has received, causing interruptions in communication between the source and destination nodes.

In the last attack scenario, the \emph{flooding attack}, the attacker node intermittently sends Route Request (RREQ) packets to randomly selected destinations within the network. The chosen destination node receives a sequence of ten consecutive RREQ broadcast messages, and this pattern repeats every 3 seconds throughout the simulation. Each of these attacks are applied for throughout the simulation (=1800 seconds).

\textbf{\emph{Dataset: }} In order to create an attack dataset, ten unique network topologies with varying network and mobility patterns are generated. Initially, all these networks are simulated without any attack. Subsequently, malicious UAVs are introduced to simulate diverse attack scenarios across the same ten topologies. The attacker ratio varies across simulations, including 5\%, 10\%, 15\%, 20\%, and 25\%. Each network simulation runs for 1800 seconds.  

\begin{table}[h!]
\caption{Dataset Information}
\label{table:dataset}
\centering
\resizebox{.6\columnwidth}{!}{\begin{tabular}{ll} 
 \toprule
Attributes & Values \\ 
 \midrule
 \# of network simulations & 160 \\ 
  \# of nodes in each network sim. & 51\\
 \# of features & 31\\
 Labels & Benign, Malicious\\
 Attack Type & Sinkhole, Blackhole, Flooding\\
 Data collection period & every 5s\\
  \bottomrule
\end{tabular}}
\end{table} 

Data is collected from each UAV periodically, every 5 seconds. Simulations were performed in a controlled environment where UAVs operated under controlled conditions. Both benign and malicious traffic were systematically introduced during these simulations. This method enabled the creation of a labeled dataset containing ground truth information, essential for developing and validating the Intrusion Detection System (IDS).  While network communications generated by the attacker nodes are labelled as malicious, the others are benign. The 31 features proposed in \cite{sen2011evolutionary} are locally collected by each UAV in this study. These features cover a range of characteristics, including aspects related to mobility, AODV control messages, and data packets \cite{sen2011evolutionary}. Examples of features include metrics providing insights into the frequency of forwarding, sending, and reception of routing protocol control packets (RREQ, RREP, and RERR). Additionally, features offer information about changes in the routing table. 

The features used for training the IDS model must be kept private. Exposure of these features compromises network confidentiality, allowing attackers to gain insights into network operations and potentially control critical communication channels. Attackers can reveal sensitive details such as network topology, traffic patterns, and node interactions, identify key nodes, and specifically target them to disrupt communication.

The dataset details are summarized in Table \ref{table:dataset}. In current study, we employed a newly adapted version of this dataset, utilizing only 10\% of the data for using in few-shot learning. The original dataset is reduced during the preprocessing stage by acquiring data collected every 50 seconds. The sizes of individual UAV dataset have been significantly reduced compared to the original dataset, consistently yielding 36 samples for each class (malicious and benign). Subsequently, a subset of these 36 samples is randomly reduced to 10 and 20 samples for utilization in other experiments.

\textbf{\emph{Methods: }}Three distinct IDSs are employed for evaluating their efficacy in detecting attacks, and these IDSs are distributed to each UAV to ensure effective operation with the data. 

The traditional \textbf{Central IDS (C-IDS)} functions by collecting data, specifically local features, from all nodes within the network. Then, it employs all features collected from UAVs to train a centralized model. The model is deployed centrally at GBS. Consequently, this model imposes a substantial burden on both the nodes and GBS, and the communication links within the network. Furthermore, it introduces a potential single point of failure in the central node.

In \textbf{\emph{Local IDS (L-IDS)}}, every UAV undergoes individual training using its own local data to develop a unique model. L-IDS is deployed on individual UAVs within a network to monitor and analyze local traffic for signs of malicious activities or policy violations. Unlike C-IDS, which collects and analyzes data from multiple sources in a centralized location, L-IDS operates independently on each device. It's important to emphasize that the evaluation results presented here represent the average of all L-IDS outcomes.

The \textbf{\emph{FL-IDS}} stands for Federated Learning-based Intrusion Detection System, which was employed in our previous study \cite{ceviz2023novel}. In this system, each UAV is initially assigned a default local model. UAVs then train their local models using their own data. However, instead of sending the actual data to a central server at the GBS, only the weights of the local models are shared. The central server aggregates these weights using FedAvg, updates the global model based on the aggregated weights, and sends the updated global model back to the UAVs for further training iterations. This iterative process continues until a specified epoch (=100 in this study) or convergence criteria are met. This approach generally reduces the burden on the communication links and the central server, as only model weights are transmitted rather than raw data. It also helps maintain privacy and security since sensitive data remains on the local devices and is not shared directly with the central server. 

The \textbf{\emph{FSFL-IDS}} stands for Federated and Few-Shot Learning-based Intrusion Detection System. This new approach combines the advantages of FL, which is well-suited for the distributed architecture of FANETs, with the benefits of few-shot learning, a technique used for training models with limited data. In FSFL-IDS, FANETs limit the communication rounds between UAVs and server to 10 in this study. This constraint helps reduce the overhead on communication links and ensures that the learning process is efficient and practical within the dynamic and resource-constrained environment of FANETs. By leveraging FL, FSFL-IDS allows nodes to train local models using their own data without the need to share raw data with a central server. Instead, only model updates or aggregated information is exchanged, preserving data privacy and security. 

Furthermore, by incorporating FSL techniques, FSFL-IDS enables models to learn effectively even when training data is limited, which is often the case in intrusion detection scenarios where obtaining labeled data can be challenging. Moreover, the challenges can intensify in dynamic settings, where network conditions fluctuate and intrusion patterns evolve rapidly. Moreover, in such scenarios, there may be instances where intrusion detection systems struggle to communicate due to unreliable or lossy links.  

All proposed frameworks apart from the L-IDS, the model is trained offline on a powerful server. UAVs communicate with the central server at predefined intervals or when connectivity is available. These intervals can be scheduled based on operational needs or network availability, such as once every few hours or when UAVs return to a base station. During these communication, each UAV uploads its local model weights or gradients to the central server. By restricting data transfer to these designated periods, the system efficiently utilizes available bandwidth and conserves the UAVs' battery life. Once the model is trained offline, it is deployed to the GBS for testing.

\textbf{\emph{Performance Metrics: }} In this study, well-known metrics, namely accuracy, recall or detection rate (DR), and false positive rate (FPR), are employed to assess the effectiveness of detection models.

\begin{table}[ht!]
\caption{Comparison of DNN and CNN for Intrusion Detection}
\label{tablel:FSFL-classifier}
\centering
\resizebox{.8\columnwidth}{!}{
\begin{tabular}{|c|c|lll|lll|}
\hline
\multicolumn{1}{|l|}{}            & \multicolumn{1}{l|}{}                  & \multicolumn{3}{c|}{DNN}                                                & \multicolumn{3}{c|}{CNN}                                                \\ \hline
\multicolumn{1}{|l|}{Attack Type} & \multicolumn{1}{l|}{Attack Ratio(\%)} & \multicolumn{1}{l|}{Accuracy} & \multicolumn{1}{l|}{DR}       & FPR     & \multicolumn{1}{l|}{Accuracy} & \multicolumn{1}{l|}{DR}       & FPR     \\ \hline
\multirow{5}{*}{Sinkhole}         & 5                                      & \multicolumn{1}{l|}{73.33\%}  & \multicolumn{1}{l|}{81.67\%}  & 18.33\% & \multicolumn{1}{l|}{75.83\%}  & \multicolumn{1}{l|}{88.00\%}  & 12.00\% \\ \cline{2-8} 
                                  & 10                                     & \multicolumn{1}{l|}{81.67\%}  & \multicolumn{1}{l|}{90.00\%}  & 10.00\% & \multicolumn{1}{l|}{90.00\%}  & \multicolumn{1}{l|}{93.33\%}  & 6.67\%  \\ \cline{2-8} 
                                  & 15                                     & \multicolumn{1}{l|}{85.83\%}  & \multicolumn{1}{l|}{73.33\%}  & 26.67\% & \multicolumn{1}{l|}{94.17\%}  & \multicolumn{1}{l|}{93.33\%}  & 6.67\%  \\ \cline{2-8} 
                                  & 20                                     & \multicolumn{1}{l|}{91.67\%}  & \multicolumn{1}{l|}{93.33\%}  & 6.67\%  & \multicolumn{1}{l|}{98.33}  & \multicolumn{1}{l|}{97.00\%}  & 3.00\%  \\ \cline{2-8} 
                                  & 25                                     & \multicolumn{1}{l|}{90.00\%}  & \multicolumn{1}{l|}{90.00\%}  & 10.00\% & \multicolumn{1}{l|}{99.17\%}  & \multicolumn{1}{l|}{98.33\%}  & 1.67\%  \\ \hline
\multirow{5}{*}{Blackhole}        & 5                                      & \multicolumn{1}{l|}{65.83\%}  & \multicolumn{1}{l|}{80.00\%}  & 20.00\% & \multicolumn{1}{l|}{77.29}  & \multicolumn{1}{l|}{62.00\%}  & 38.00\% \\ \cline{2-8} 
                                  & 10                                     & \multicolumn{1}{l|}{83.33\%}  & \multicolumn{1}{l|}{78.33\%}  & 21.67\% & \multicolumn{1}{l|}{91.67\%}  & \multicolumn{1}{l|}{91.67\%}  & 8.33\%  \\ \cline{2-8} 
                                  & 15                                     & \multicolumn{1}{l|}{90.42\%}  & \multicolumn{1}{l|}{86.67\%}  & 13.33\% & \multicolumn{1}{l|}{95.00\%}  & \multicolumn{1}{l|}{93.33\%}  & 6.67\%  \\ \cline{2-8} 
                                  & 20                                     & \multicolumn{1}{l|}{88.33\%}  & \multicolumn{1}{l|}{78.33\%}  & 21.67\% & \multicolumn{1}{l|}{98.33\%}  & \multicolumn{1}{l|}{98.33\%}  & 1.67\%  \\ \cline{2-8} 
                                  & 25                                     & \multicolumn{1}{l|}{93.33\%}  & \multicolumn{1}{l|}{90.00\%}  & 10.00\% & \multicolumn{1}{l|}{99.17\%}  & \multicolumn{1}{l|}{98.33\%}  & 1.67\%  \\ \hline
\multirow{5}{*}{Flooding}         & 5                                      & \multicolumn{1}{l|}{71.46\%}  & \multicolumn{1}{l|}{63.33\%}  & 36.67\% & \multicolumn{1}{l|}{78.33\%}  & \multicolumn{1}{l|}{86.67\%}  & 13.33\% \\ \cline{2-8} 
                                  & 10                                     & \multicolumn{1}{l|}{84.43\%}  & \multicolumn{1}{l|}{70.00\%}  & 30.00\% & \multicolumn{1}{l|}{95.26\%}  & \multicolumn{1}{l|}{98.00\%}  & 2.00\%  \\ \cline{2-8} 
                                  & 15                                     & \multicolumn{1}{l|}{88.33\%}  & \multicolumn{1}{l|}{100\%} & 0\%  & \multicolumn{1}{l|}{99.17\%}  & \multicolumn{1}{l|}{100\%} & 0\%  \\ \cline{2-8} 
                                  & 20                                     & \multicolumn{1}{l|}{90.00\%}  & \multicolumn{1}{l|}{100\%} & 0\%  & \multicolumn{1}{l|}{100\%}    & \multicolumn{1}{l|}{100\%} & 0\%  \\ \cline{2-8} 
                                  & 25                                     & \multicolumn{1}{l|}{99.17\%}  & \multicolumn{1}{l|}{98.33\%}  & 1.67\%  & \multicolumn{1}{l|}{100\%}    & \multicolumn{1}{l|}{100\%} & 0\%  \\ \hline
\end{tabular}
}
\end{table}

\section{Experimental Results}
In this section, we present the outcomes of our experiments, conducting a thorough evaluation of FSFL-IDS. 
We evaluate and compare FSFL-IDS, FL-IDS, and L-IDS utilizing a downscaled dataset by 10\% across sinkhole, blackhole, and flooding attacks under varying attacker ratios in FANETs. Moreover, we present the impact of the number of samples (10 and 20 shots) on intrusion detection performance. In this context, the performance of both traditional C-IDS and FSFL-IDS is evaluated. All IDS models are trained offline on a powerful computational device and subsequently deployed to the ground base station (GBS) for the testing process.

\subsection{Classifier: DNN vs CNN}

We firstly compared DNN and CNN algorithms' performance on various simulated attack types using FSFL-IDS, as detailed in Table \ref{tablel:FSFL-classifier}. CNN consistently outperforms DNN in accuracy and DR across all scenarios. The performance gap widens for sinkhole and blackhole attacks, especially at the 25\% attacker ratio. Even at the lowest attacker ratio, CNN outperforms DNN, thanks to its inherent complexity enabling more effective classification with less data. Additionally, CNN exhibits a lower FPR compared to DNN, enhancing reliability and efficiency by reducing false alarms for benign activities in intrusion detection systems.

\begin{figure}[h]
  \centering
  \includegraphics[width=.8\textwidth]{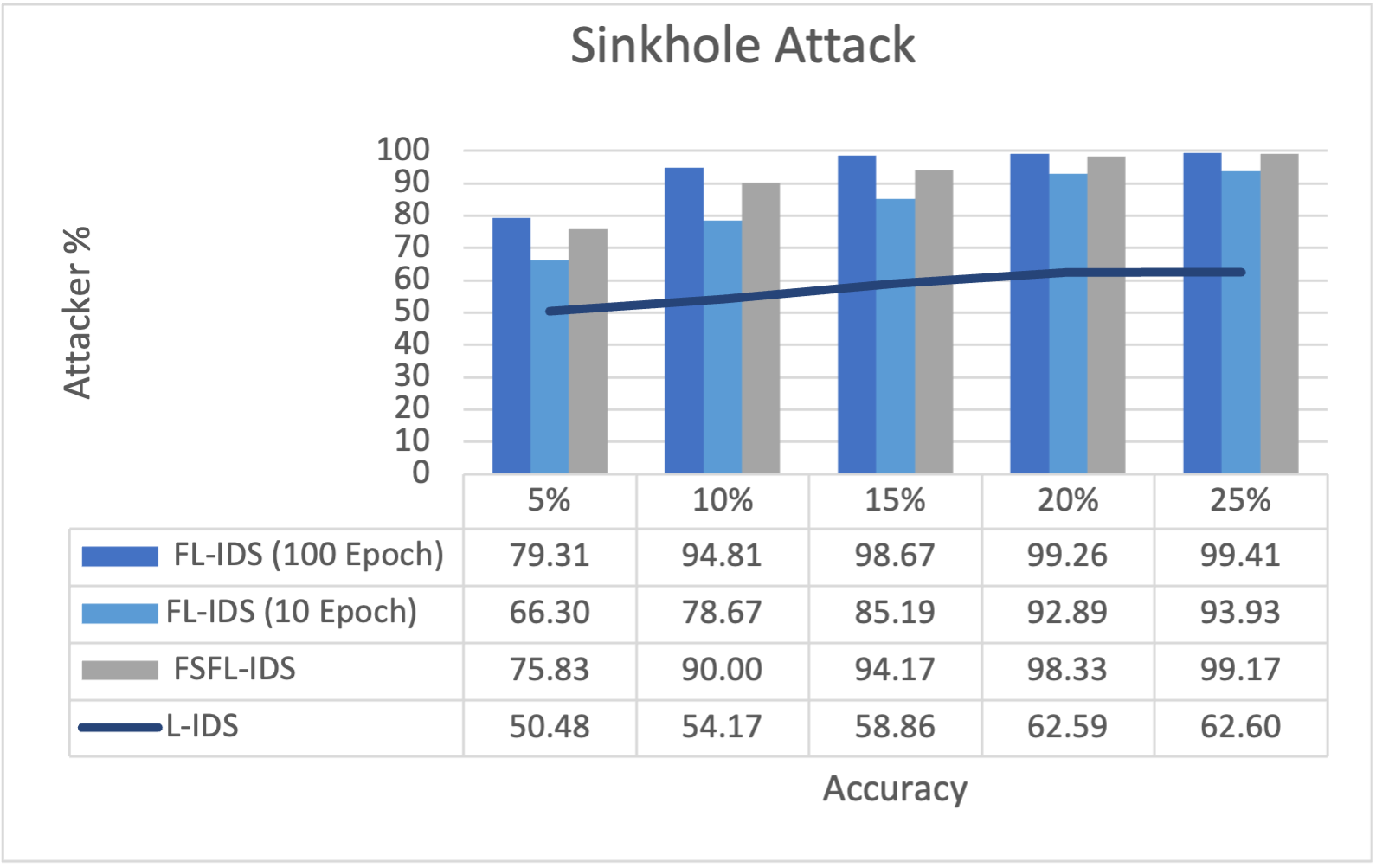}
  \caption{Comparison of FL-IDS and FSFL-IDS in Detecting
Sinkhole Attack}
  \label{sh-accuracy}
\end{figure}

\subsection{Effectiveness: L-IDS, FSFL-IDS and FL-IDS}

In this section, we compare the detection capabilities of L-IDS, FL-IDS and FSFL-IDS. Both systems are optimized using the Hyperband algorithm to ensure a fair comparison. This comprehensive examination provides insights into the performance of each IDS type across various attack types and network conditions. 




\textbf{\emph{Sinkhole Attack: }}
The results from both IDSs regarding sinkhole attack detection are depicted comprehensively in Figure \ref{sh-accuracy}. Sinkhole attacks, typically used as initial attacks, passively redirect network traffic and are less aggressive compared to blackhole and flooding attacks. Their focus solely on traffic redirection suggests a lesser impact on the network, particularly at lower attacker ratios. However, in FSFL-IDS, reducing the number of samples at lower attacker ratios may compromise detection efficacy. Notably, FL-IDS (79.31\%) outperforms FSFL-IDS (75.83\%) at 5\% attacker ratio. On the other hand, L-IDS achieves the lowest accuracy of 50.48\% with a significant gap compared to the other IDSs at 5\% attacker ratio. Yet, as the attacker ratio increases to 20\% or higher, the performance gap between FL-IDS and FSFL-IDS decreases to less than 1\% due to increased sinkhole attack impact on nodes while it increases for L-IDS. While FL-IDS consistently produces the lowest results across all ratios at 10 epochs, it is observed that its performance improves when the attacker ratio exceeds 15\%. However, to achieve acceptable results even at lower ratios, it requires 100 epochs, thereby leading to increased communication costs.
In summary, even with only 10\% of the samples used in training, FSFL-IDS demonstrates comparable performance to FL-IDS, highlighting the potential advantages of the FSFL-IDS approach, especially in resource-constrained and distributed environments, even for less aggressive and challenging-to-detect attack scenarios.

\begin{figure}[h]
  \centering
  \includegraphics[width=.8\textwidth]{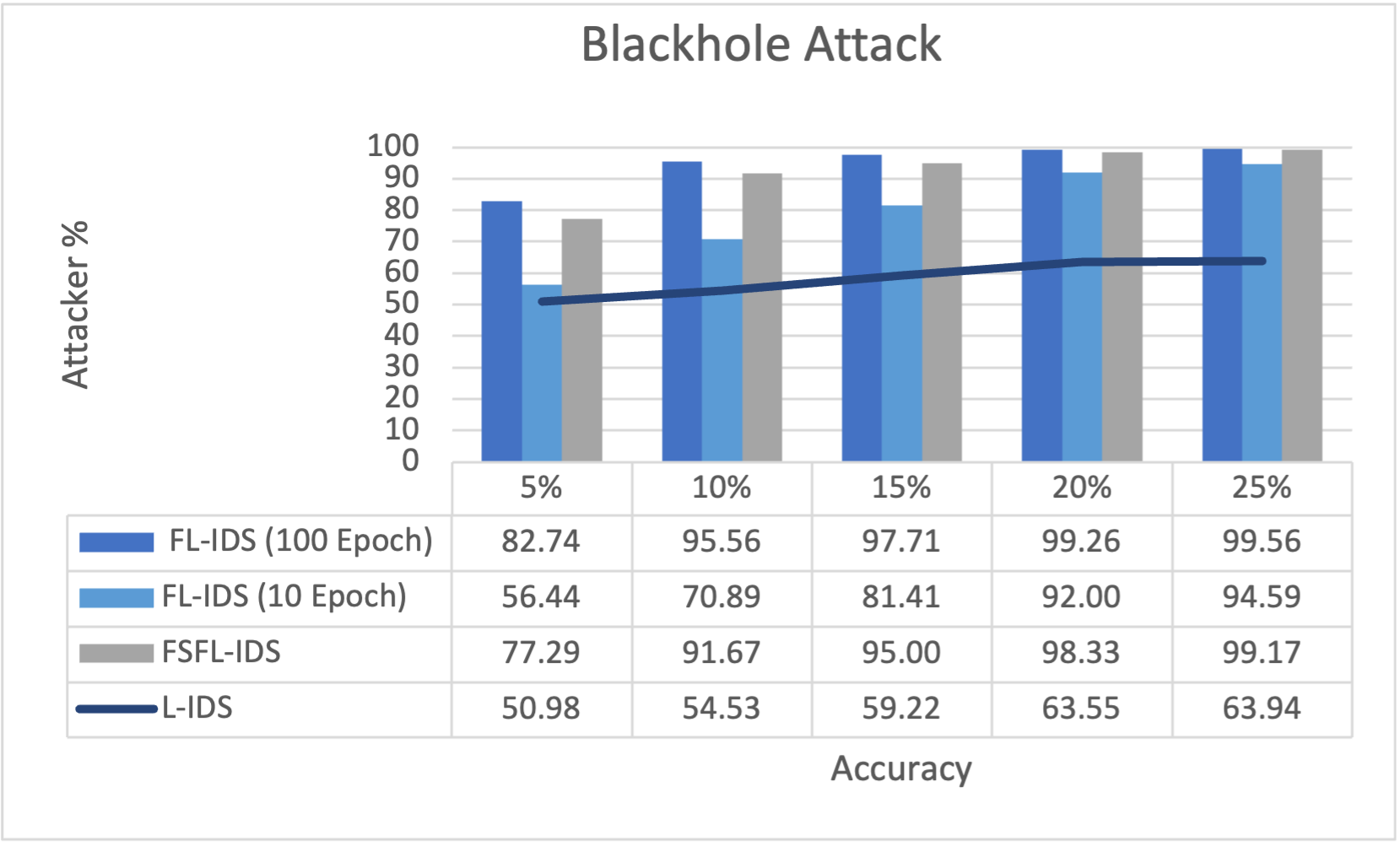}
  \caption{Comparison of FL-IDS and FSFL-IDS in Detecting
Blackhole Attack}
  \label{bh-accuracy}
\end{figure}

\textbf{\emph{Blackhole Attack: }}The results for detecting blackhole attacks across the three IDSs are comprehensively illustrated in Figure \ref{bh-accuracy}. Initially, at a low attacker ratio of 5\%, FL-IDS (82.74\%) outperforms FSFL-IDS (77.29\%) and L-IDS (50.98\%) in detecting blackhole attacks. 
The uneven influence of attacker nodes because of their varied positions within the network leads to disparate data distribution, resulting low detection  performance for FSFL-IDS at low attacker ratios. However, beyond 15\% attacker ratio, FSFL-IDS narrows the performance gap with FL-IDS.  Specifically, at 25\% attacker ratio, the accuracy difference between FSFL-IDS and FL-IDS reduces to less than 0.4\%. This difference indicates that FSFL-IDS, which processes a limited amount of data with fewer training epochs, can achieve comparable results with FL-IDS at higher attacker ratios with less computational and communication cost. 
On the other hand, even with the highest attacker ratio, L-IDS still shows the lowest accuracy results with nearly a 30\% difference from other IDSs.

Nevertheless, as the attacker ratio in the network increases, there is a notable enhancement in accuracy across all methodologies. This is attributed to the increased presence of more attacker nodes in the network, providing deep learning algorithms with additional anomaly data and patterns for model training. Consequently, the performance of FL-IDS and FSFL-IDS approaches improves with the increase in the attacker ratio. As expected, the peak performance for FSFL-IDS (99.17\%) is attained at the highest attacker ratio.

\begin{figure}[h]
  \centering
  \includegraphics[width=.8\textwidth]{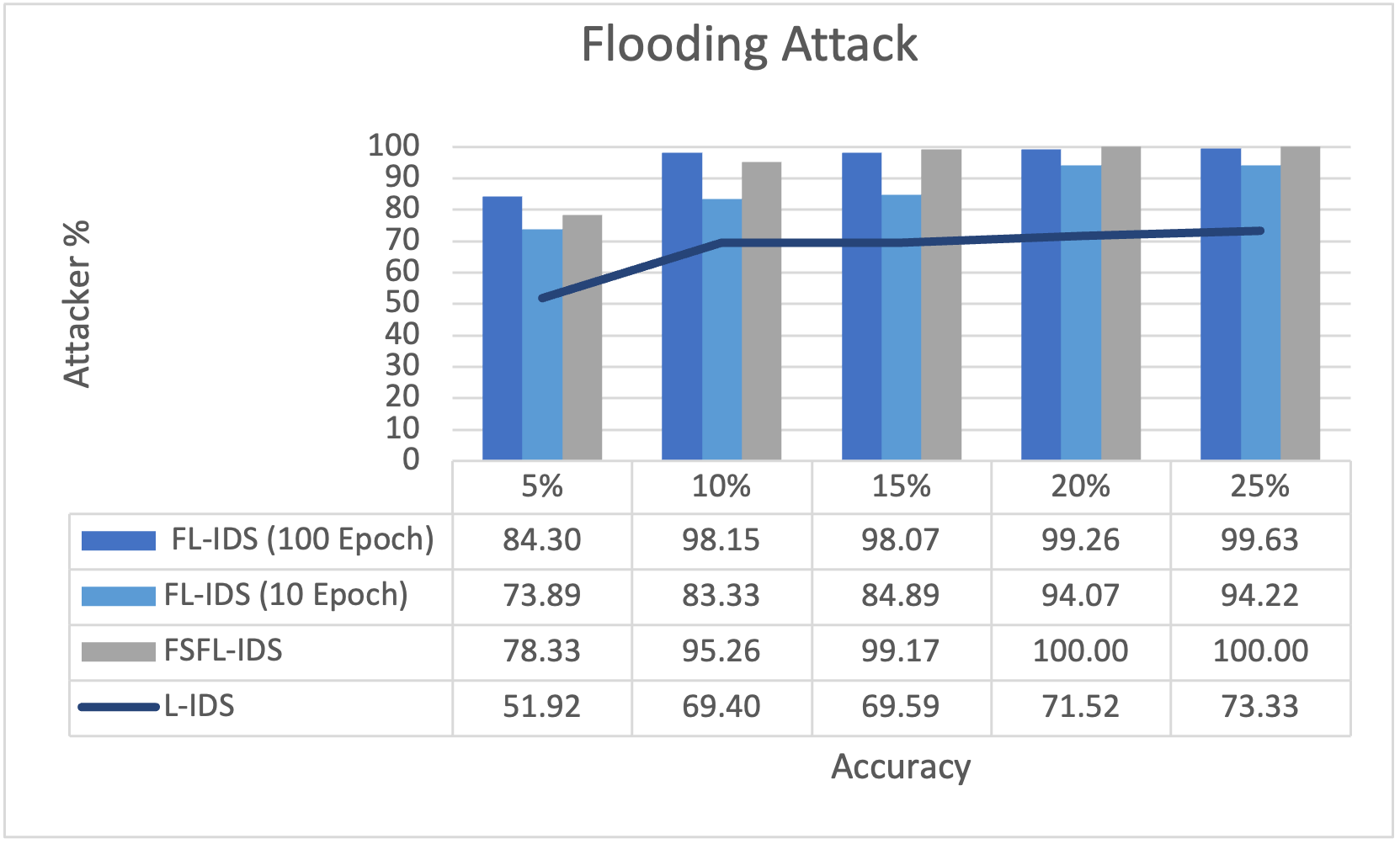}
  \caption{Comparison of FL-IDS and FSFL-IDS in Detecting
Flooding Attack}
  \label{fl-accuracy}
\end{figure}

\textbf{\emph{Flooding Attack: }} 
Figure \ref{fl-accuracy} compares three IDS approaches in detecting flooding attacks. Unlike other attacks, flooding attacks achieve high accuracy even with a low attacker ratio, as low as 10\%, attributed to their nature. All IDSs achieve accuracy exceeding 98\%, with FSFL-IDS slightly outperforming FL-IDS, especially at 15\% attacker ratio and beyond except L-IDS. L-IDS achieves better performance in detecting flooding attacks compared to other attacks. However, its performance is still significantly lower than that of other methods. Flooding attacks, affecting the entire network, can be detected successfully with a smaller dataset, advantageous for discerning subtle differences. FSFL-IDS emerges as the more suitable choice, given the pronounced distinguishing features of flooding attacks.

\begin{table}[]
\caption{Evaluation Results of C-IDS and FSFL-IDS in terms of Sample Numbers}
\label{tab:comp-sample}
\resizebox{\columnwidth}{!}{\begin{tabular}{|c|cll|ll|ll|}
\hline
 &
  \multicolumn{3}{c|}{Sinkhole} &
  \multicolumn{2}{c|}{Blackhole} &
  \multicolumn{2}{c|}{Flooding} \\ \hline
 &
  \multicolumn{1}{c|}{Model} &
  \multicolumn{1}{c|}{10\% attacker rt.} &
  \multicolumn{1}{c|}{\%25 attacker rt.} &
  \multicolumn{1}{c|}{\%10 attacker rt.} &
  \multicolumn{1}{c|}{\%25 attacker rt.} &
  \multicolumn{1}{c|}{\%10 attacker rt.} &
  \multicolumn{1}{c|}{\%25 attacker rt.} \\ \hline
\multirow{2}{*}{10-shot} &
  \multicolumn{1}{l|}{C-IDS} &
  \multicolumn{1}{l|}{63.80\%} &
  81.32\% &
  \multicolumn{1}{l|}{64.67\%} &
  84.93\% &
  \multicolumn{1}{l|}{93.50\%} &
  95.18\% \\ \cline{2-8} 
 &
  \multicolumn{1}{c|}{FSFL-IDS} &
  \multicolumn{1}{l|}{62.07\%} &
  64.70\% &
  \multicolumn{1}{l|}{63.03\%} &
  68.15\% &
  \multicolumn{1}{l|}{91.38\%} &
  94.12\% \\ \hline
\multirow{2}{*}{20-shot} &
  \multicolumn{1}{l|}{\begin{tabular}[c]{@{}l@{}}C-IDS \end{tabular}} &
  \multicolumn{1}{l|}{85.28\%} &
  95.42\% &
  \multicolumn{1}{l|}{88.93\%} &
  96.42\% &
  \multicolumn{1}{l|}{97.22\%} &
  99.25\% \\ \cline{2-8} 
 &
  \multicolumn{1}{c|}{FSFL-IDS} &
  \multicolumn{1}{l|}{83.15\%} &
  93.87\% &
  \multicolumn{1}{l|}{84.08\%} &
  94.76\% &
  \multicolumn{1}{l|}{93.09\%} &
  99.17\% \\ \hline
\end{tabular}}
\end{table}

\textbf{\emph{General Discussion: }}

In scenarios involving flooding, sinkhole, and blackhole attacks, both at lower and higher attacker ratios, L-IDS consistently demonstrated lower accuracy compared to FL-IDS and FSFL-IDS. Only in the case of flooding attacks does L-IDS show relatively better results at higher attacker ratios, as this type of attack tends to impact a majority of the nodes within the network. Considering the results, L-IDS is not suitable for dynamic distributed systems due to its consistently lower accuracy. L-IDS may have limitations in detection accuracy due to the restricted view of network activities and the lack of collaborative data analysis across the network, making it less effective in dynamic and distributed systems. 

Utilizing training models with extensive data and high epochs enhances the ability to differentiate attacks from benign network traffic. Consequently, FL-IDS consistently outperforms across various attack types at 100th epoch, while FSFL-IDS competes closely with FL-IDS, especially at higher attacker ratios. However, for all attack types, FL-IDS consistently shows lower results at the 10th epoch comparing to the FSFL-IDS. FL-IDS demonstrates effective performance for scenarios with high attacker ratios at 10 epochs. However, it is inadequate for scenarios with low attacker ratios.

However, in scenarios where the attack impact is limited, such as with sinkhole and blackhole attacks affecting specific victims and their surroundings, FSFL-IDS may exhibit lower average accuracy at low attacker ratios. Moreover, detecting a blackhole attack is generally easier than detecting a sinkhole attack due to its more active pattern of dropping data packets.

In the case of flooding attacks, even at lower attacker ratios, all nodes experience significant impact due to the attack's ubiquitous nature. As a result, FSFL-IDS demonstrates high detection accuracy even at lower attacker ratios. This positive contribution is attributed to FSFL-IDS's ability to learn the general pattern with limited data when more nodes are impacted by such attacks.

In summary, while FL-IDS outperforms FSFL-IDS at lower attacker ratios, particularly in networks with a significant number of attackers (20\%, 25\%), the performance of FSFL-IDS approaches close to an ideal level. Even at low attacker ratio (10\%), flooding attacks drive IDS convergence toward the optimal due to their widespread impact on nearly all nodes.

\subsection{Efficiency: FSFL-IDS vs FL-IDS} 
As shown above, FL-IDS exhibits a high performance in detecting various network attacks,  such as blackhole, sinkhole, and flooding attacks. Particularly it outperforms FSFL-IDS particularly at lower attacker ratios. On the other hand, FSFL-IDS reduces communication, computation and storage costs.

\textbf{\emph{Communication Cost: }}
The communication cost (CC) of FL-IDS or FSFL-IDS can be computed using the following formula: 
\begin{equation}
CC = N * W * E * S 
\end{equation}
where \emph{N} represents the total number of nodes, \emph{W} is the number of weights sent per communication round, \emph{E} is the number of communication rounds, and \emph{S} is the average weight size. By reducing the communication rounds to 10 in FSFL-IDS compared to 100 in FL-IDS, FSFL-IDS achieves a 10\% reduction in communication cost. This reduction is crucial, especially for energy-constrained devices like mini UAVs, and advantageous in dynamic environments with lossy links.  In larger scaled FANETs with more UAVs, minimizing communication rounds is paramount to reducing bandwidth consumption and latency in dynamic network environments characterized by variable node connectivity and mobility patterns. The dynamic nature of FANETs often leads to connectivity interruptions, requiring rapid detection with learning from limited labeled examples. This facilitates swift model updates.

\textbf{\emph{Computational and Storage Cost: }}
In FSFL-IDS, the reduction in the number of rounds, coupled with a smaller dataset length, leads to reduced training time and significantly lower computational costs per node compared to FL-IDS. FSFL-IDS utilizes only 10\% of the training data used by FL-IDS at each client. This also holds true for the Global Server (GBS), since the reduction in communication rounds further decreases computational costs, resulting in reduced training time for both clients and the GBS. Additionally, a shorter dataset length not only implies less training time but also directly influences storage costs. 

Moreover, the time-consuming nature of model training with a huge dataset poses challenges for providing a prompt response to emerging attacks. Furthermore, collecting limited data on clients helps prevent substantial information leakage in the event of a potential breach, thereby preserving the system's confidentiality and privacy. Optimizing decentralized model aggregation and local training processes is essential to maximize computational efficiency and minimize processing delays, thereby supporting scalable deployment in energy-sensitive large FANET environments.

To sum up, since the demand for high energy consumption during data acquisition, storage, and processing poses significant challenges for resource-constrained devices, affecting both clients and the global server, FSFL-IDS can be a better alternative for such environments.

\subsection{Evaluation: C-IDS and FSFL-IDS}

In this experiment, we employed N-way K-shot selection, specifically configuring cases for 2-way 10-shot and 2-way 20-shot scenarios. Due to the presence of normal and attack categories in each dataset, we limited the classification to 2-way (benign and attack). The terms "10-shot" and "20-shot" denote the use of ten and twenty samples, respectively, for learning in each category during training. Therefore, each training task encompasses N*K samples.

The objective of these experiments is to assess the practical implementability of the FSFL-IDS model in real-world scenarios. This is particularly relevant because UAVs, with their resource constraints, aim to minimize unnecessary resource utilization. Additionally, the network faces the risk of data losses due to high mobility. Therefore, training or making decisions with limited data (10 and 20) becomes crucial. The experimental results, presented in Table \ref{tab:comp-sample}, demonstrate the test accuracy of the trained models after the fine-tuning process. To evaluate the performance of the IDS based on varying sample sizes at low and high attacker ratios, we specifically employ 10\% and 25\% attacker ratios. In this comparative experiment, we also evaluated the central IDS deployed on GBS.

The results indicate that, for FSFL-IDS to achieve reasonable accuracy, at least 20 shots need to be used for local training and shared with the global model. Particularly for blackhole and sinkhole attacks, the 10-shot FSFL-IDS struggles to distinguish malicious traffic from benign traffic effectively. In contrast, due to the ubiquitous nature of flooding attacks, high accuracy can be achieved even with only 10 shots.

In contrast, C-IDS can yield satisfactory results even with 10-shot approach. In this setup, local data comprising 10 or 20 samples for each class is periodically transmitted from each client to the central IDS, which directly employs them as inputs for training the global model. Despite the limited number of shots used for training in the global model, the volume of data collected from each client increases significantly. As a result, C-IDS achieves higher accuracy compared to FSFL-IDS. Furthermore, as both the attacker ratio and the number of samples increase, C-IDS consistently improves its performance, whereas FSFL-IDS produces comparable results. This observation can be attributed to C-IDS benefiting from access to the entire dataset generated by the network. 
This comprehensive network view significantly contributes to its superior accuracy in detecting attacks, albeit with an increased communication cost and privacy issues.

\subsection{Limitations}
FSFL-IDS, designed to enhance IDS in FANETs, faces several critical limitations that impact its effectiveness and reliability in dynamic network environments. One significant challenge is the heterogeneity of data across nodes within FANETs, where variations in data quality and availability can hinder the convergence and accuracy of federated learning processes. This diversity in data sources may lead to uneven contributions to the global IDS model, potentially compromising detection accuracy. Operational constraints, including frequent topology changes and intermittent connectivity, further challenge FSFL-IDS by disrupting consistent model synchronization and timely data updates. Moreover, ensuring robust security against adversarial attacks, such as model poisoning, is critical. Adversaries could exploit vulnerabilities in the aggregation process to inject malicious updates, compromising the integrity and efficacy of the IDS. Addressing these challenges is essential for deploying FSFL-IDS effectively in real-world UAV scenarios, ensuring both security and performance in dynamic and heterogeneous environments.

\section{Conclusion}
Given the resource-intensive nature of model retraining, the significance of data privacy, whether on the wire or within the device, and the challenges posed by the dynamic network environment in FANETs, which involves topology changes and potential data loss used for model training or detection, it is crucial to explore alternative security solutions. Our implementation, incorporating few-shot learning and federated learning, inherently minimizes the risk caused by data loss. In the federated learning process, the model is trained within the nodes, and only the model weights are transmitted for aggregation, eliminating the need to send raw data. This strategic approach ensures privacy and mitigates concerns related to data loss in dynamic network environments.

We can further assess the feasibility of FSFL-IDS, taking into account its reduced communication and computation costs while leveraging the privacy consideration of federated learning. As discussed earlier, FSFL-IDS presents a promising avenue, demonstrating competitive accuracy levels while significantly mitigating the burden on resources. This strategic evaluation becomes particularly crucial for devices with limitations, relying on low-capacity batteries. Our study further underscores the strengths and considerations of FL-IDS and FSFL-IDS across various attack types. FSFL-IDS could detect simulated attacks with high accuracy, particularly at higher attacker ratios, demonstrating commendable performance against blackhole, sinkhole, and flooding attacks.

By adopting the N-way K-shot approach, specifically in 2-way 10-shot and 2-way 20-shot settings, we aimed to mimic real-world scenarios, 
where data losses might occur due to connection disruptions. The results showed FSFL-IDS's resilience and competitive accuracy, showing comparable performance to C-IDS.

In summary, our study navigates the complexities of intrusion detection in dynamic networks, providing valuable and promising insights into the strengths and challenges of FL-IDS and the promising adaptability of FSFL-IDS. While FANETs are shown as exemplar networks here, we believe the insights gained and methodologies developed in this study hold promise for application in a wide range of dynamic network environments such as Internet of Things (IoT), other type of mobile ad hoc networks and sensor networks.

%
%
%
 \bibliographystyle{splncs04}
\bibliography{few-shot-securecomm}

\end{document}